\documentclass{osa-article}

\journal{osac}

\articletype{Research Article}

\begin{document}

\title{Efficient Solar-driven Steam Generation Enabled by An Ultra-black Paint}

\author{Xiaojie Liu,\authormark{1} Yanpei Tian,\authormark{1} Fangqi Chen,\authormark{1} Ralph Ahlgren,\authormark{2} Yiting Zheng,\authormark{3} Ming Su,\authormark{3} Gang Xiao,\authormark{4} and Yi Zheng\authormark{1,*}}

\address{\authormark{1}Department of Mechanical and Industrial Engineering, Northeastern University, Boston, MA 02115, USA\\
\authormark{2}Soleeva Energy, Inc. San Jose, CA 95131, USA\\
\authormark{3}Department of Chemical Engineering, Northeastern University, Boston, MA 02115, USA\\
\authormark{4}Department of Physics, Brown University, Providence, RI 02912, USA}

\email{\authormark{*}opex@osa.org} 



\begin{abstract}
 Solar-driven interfacial steam generation for desalination has attracted broad attention. However, a significant challenge still remains for achieving a higher evaporation rate and high water quality, together with a cost-effective and easy-to-manufacture device to provide a feasible solar-driven steam generation system. In this study, a novel ultra-black paint, Black 3.0, serving as a perfect solar absorber is introduced into the hot-pressed melamine foam networks, allowing us to construct an ultra-black (99\% absorptance in the solar region) and self-floating evaporation device. The high performing features of effective solar absorptance and salt-rejection capability contribute to a high-to-date evaporation rate of freshwater at 2.48 kg m$^{-2}$ h$^{-1}$ under one sun (1 kW m$^{-2}$). This interfacial solar evaporator has a daily drinkable water yield of 2.8 kg m$^{-2}$ even in cloudy winter weather and maintains stability in water with a wide range of acidity and alkalinity (pH 1\textasciitilde14). These features will enable the construction of a facilely fabricated, robust, highly-efficient, and cost-effective solar steam generation system for freshwater production.
\end{abstract}

\section{Introduction}
Access to freshwater is of pivotal importance to humanities. Fast population growth and climate change demand increasing supply of freshwater. Water scarcity has become a threat to the sustainable development of human society \cite{mekonnen2016four,reif2015solar}. This motivates the development of utilizing saline waters from the oceans and other brackish water sources and the processes that convert saltwater into freshwater. Solar energy is now emerging as one of the most promising sustainable energy sources, as it is clean and can be supplied without any environmental pollution compared with other forms of energy  \cite{kalogirou2004solar}. And the abundant solar energy makes the solar-driven evaporation one of the promising approaches for water desalination and purification. 

In a conventional solar-driven evaporation system, bulk water is heated to a high temperature to generate water vapor, resulting in a slow response to sunlight and heat loss to the bulk water or the external environment \cite{kabeel2011review,neumann2012solar,tao2018solar,lenert2012optimization,hogan2014nanoparticles}. In 2011, Wang \textit{et al.,} were among the first to demonstrate the floating interfacial solar-driven evaporation structure into desalination area, which does not require significant capital investment in high-cost permanent construction and/or land use \cite{zeng2011solar}. Compared with bulk water-heating method which heats the entire body of water, the interfacial solar-driven evaporation approach mainly localizes the heat generation at the water-air interface. This method avoids heating a large volume of water, \textit{e.g.,} the ocean, which serves as a low-temperature sink. The interfacial evaporation has an ultra-fast response to sunlight and is endowed with a much higher systematic thermal efficiency \cite{wang2018emerging,wang2014bio}. Furthermore, the technique can be easily applied by floating an absorber sheet on the water surface without complex pressure control or other bulky infrastructure. Owing to all these advantages, scientists have made great efforts to develop the interfacial solar-driven evaporation.

To yield a remarkable photothermal performance, the development of interfacial evaporation structure should center around the following four key factors. (1) The ideal solar absorber should possess outstanding absorptance in the range of visible and near-infrared regions \cite{zhang2017vertically} to convert the solar radiation into heat to be used for water vapor generation; (2) It is crucial for the interfacial evaporation to feature a low thermal conductivity, thereby reducing the heat loss from the absorber to the bulk water and localizing the heat at the air/water evaporative interface; (3) High hydrophilicity and porous framework are necessary for sufficient water transport from the bulk water to the heated area; and (4) Self-floating ability is required to avoid land use and/or permanent infrastructure construction. Even with intensive efforts so far, it remains as a huge challenge to develop an approach that can satisfy these four criteria simultaneously.

To obtain a high efficiency, a multilayered structure is proposed to take into account these pivotal factors and it provides us with more options to choose alternative appropriate materials \cite{shi2017rational,lou2016bioinspired,ghasemi2014solar,ni2018salt,li2016graphene}. Ghasemi \textit{et al.,} first demonstrated the double-layered structure consisting of a carbon foam layer supporting an exfoliated graphite layer \cite{ghasemi2014solar}. The bottom layer serves as a heat barrier to minimize the heat loss to the bulk water, and its porous structure is used to transport water to the heated area, and meanwhile it supports the top layer to make it possible for self-floating. The top porous layer serves as the absorber for the vapor transport.

Recently, many similar double-layered evaporation structures have been  widely adopted. In most cases, such as air-laid paper \cite{liu2015bioinspired}, polystyrene foam \cite{shi2017rational}, woodblock \cite{kim2018mesoporous,jiang2017water,chen2017highly}, polyurethane sponge \cite{zhang2019facile}, macroporous silica substrate \cite{wang2016self}, and bacterial nanocellulose aerogel \cite{jiang2016bilayered}, the bottom layer serves both as a heat barrier and a water transport medium. Under a steady working condition, when water is transported to the top photothermal area through the bottom layer, almost all of the macropores inside the heat barrier are filled with water which typically is more thermally conductive than the heat barrier. This process renders the heat barrier less effective. To avoid heat loss, a new multilayer consisting of photothermal material, closed-pore thermal insulator, and external hydrophilic materials has generated much interest. The closed-pore thermal insulator serves as a real heat barrier layer and the air inside the insulator reduces heat conduction. For example, polystyrene foam \cite{li2016graphene,ni2016steam,li2017three,ni2018salt,xu2017mushrooms,liu2019high} is a good and inexpensive thermal insulator with low thermal conductivity. External hydrophilic materials, wrapping or passing through the insulator foam, act as a capillary-driven pump with a one-/two-dimensional water path \cite{liu2017bioinspired,li2016graphene}. By and large, this modified multilayered evaporation structure has become a favorable way to further minimize heat conduction losses and improve energy conversion efficiency. 

An efficient solar steam generation depends highly on the photothermal conversion materials with broadband absorptance and high photothermal properties \cite{zhu2018solar,gao2019solar}. On the basis of a multilayered structure, the diverse photothermal materials with wide light absorbance have been extensively studied including carbon-based materials \cite{zeng2011solar,ghasemi2014solar,hu2017tailoring,zhou2016macroscopic}, metals \cite{brongersma2015plasmon,richardson2009experimental,liu2017solar,zhou20163d}, and metal oxides \cite{zhu2016constructing,ye2017synthesis,wang2017high,ding2017non}. Carbon-based absorbers, such as carbon black \cite{liu2015floatable}, graphene \cite{yang2018graphene}, graphene oxide \cite{li2016graphene,li20173d,jiang2016bilayered} and carbon nanotube \cite{chen2017highly,wang2016self}, offer great advantages of high solar absorbance and thermal stability. They have become the largely approving choices, however, the utilization of elaborate technologies and the high-associated cost limit their field applications to large-scale desalination \cite{lin2018integrative,tao2018solar,li2016graphene}. Although the metal-based (\textit{e.g.,} gold nanoparticles) materials have attracted tremendous attention due to their unique optical and photothermal properties \cite{loeb2017solar,liu2015bioinspired,zhou2017self}, their industrial applications are limited because of their inferior chemical and thermal stabilities, exorbitant cost, and complex fabrication processes.

With the aim to further explore novel techniques and materials in improving evaporation efficiency and reducing cost for desalination, we have applied the blackest paint, called Black 3.0 Paint\textregistered, to the floating interfacial solar-driven evaporation device serving as the photothermal conversion material combined with a sheet of melamine foam (MF).  Herein, a sheet of hot-pressed MF plays an ideal elastic skeleton material sprayed with Black 3.0 paint, serving on top as a perfect solar absorber to efficiently absorb and convert the solar radiation into heat. To achieve a perfect evaporation performance and high solar-thermal energy conversion efficiency, we demonstrate a self-floating three-layer evaporation structure with a two-dimensional water path to localize solar-thermal heat generation to the air/water interface. This entire device is made from the commercially available and low-cost materials: solar absorber uses a sheet of hot-pressed MF combined with Black 3.0 paint; two-dimensional water path is provided by the highly absorbent Webril all-cotton wipes, and thermal barrier uses the Polyvinyl chloride (PVC) foam. Under 1 sun illumination without solar concentration, this evaporation device yields an excellent evaporation rate of freshwater as high as 2.48 kg m$^{-2}$ h$^{-1}$ and a highlighted evaporation efficiency of 172.5\%. Based on this multilayered evaporation structure, the novel Black 3.0 paint demonstrates a superb performance with cost-effectiveness, operability, and durability, which can be regarded as one of the indispensable choices of the solar-absorbing materials in the future development of the solar-driven evaporation process.

\begin{figure}[!t]
\centering
\includegraphics[width=0.8\textwidth]{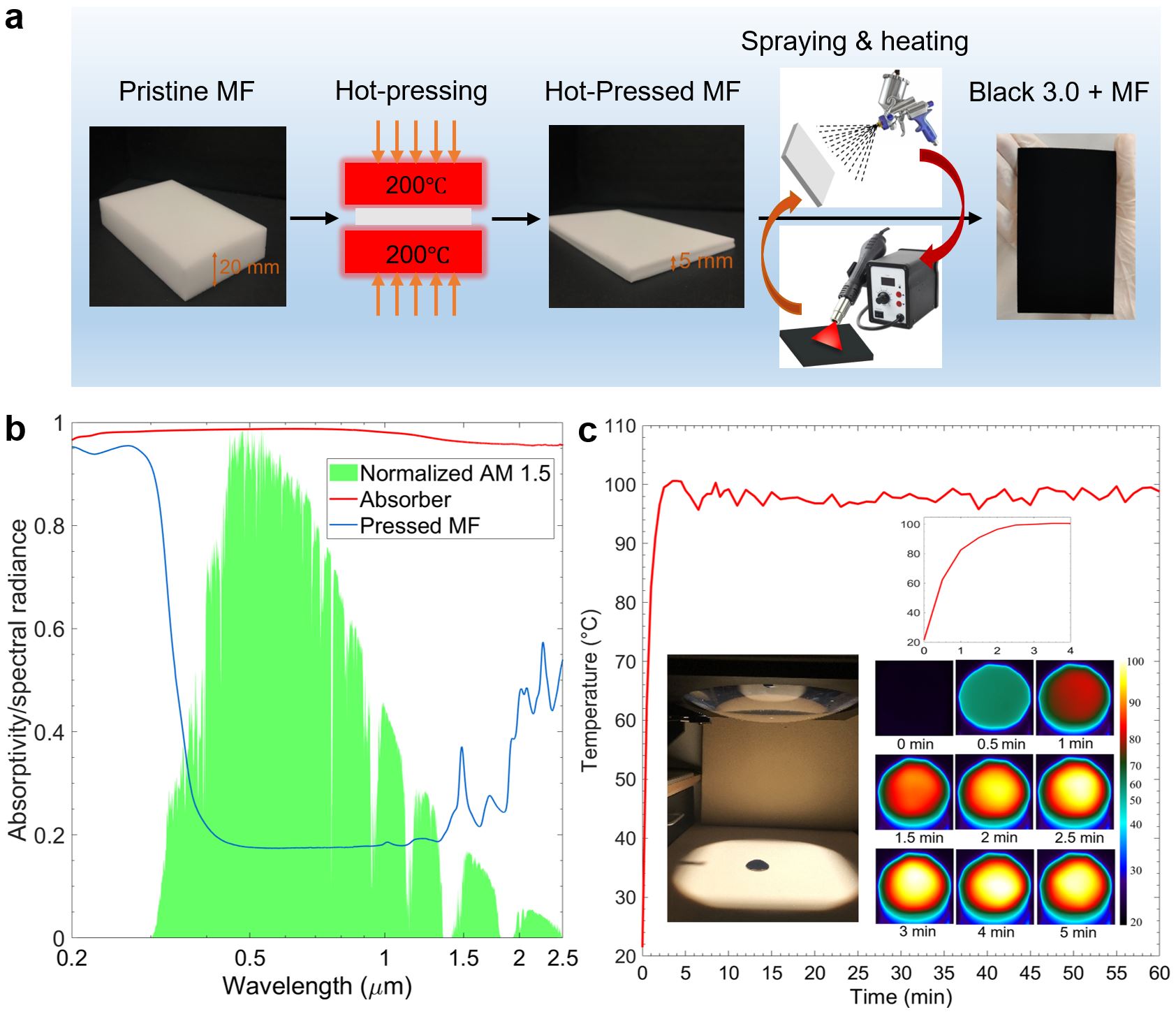}
\caption{ \label{fig:absorber} Preparation process and characterizations of the black-sprayed hot-pressed melamine foam (MF) based absorber layer. (a) Schematic illustration of the fab processes of an absorber layer in the evaporation device. (b) UV-Visible-Near-infrared absorptivity spectra of the absorber layer, the hot-pressed MF, and the normalized spectral solar irradiance density of air mass 1.5 (AM 1.5 G) global tilt solar spectrum. (c) The temperature of the absorber sheet over time under 1 sun. } 
\end{figure}

\section{Results and Discussions}

\begin{figure}[!t]
\centering
\includegraphics[width=0.8\textwidth]{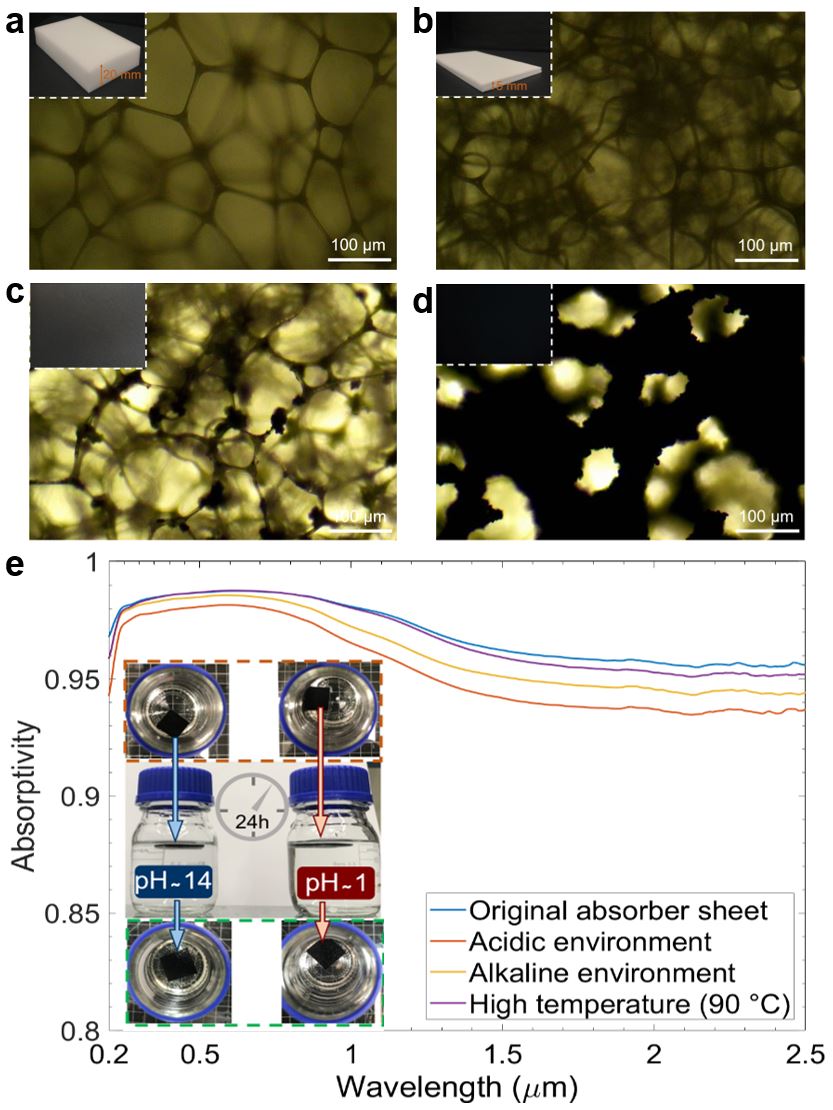}
\caption{ \label{fig:test_absorber} Microscope images of (a) pristine MF, (b) pressed MF, (c) one-layer paint coated MF, and (d) three-layer paint coated MF (a-d, insets are the photographs of the pristine/hot-pressed, one-/three-layer paint coated MFs). (e) The UV-Visible-Near-infrared absorptivity spectra measurement and the chemical stability tests of the absorber layer under different harsh environments. The inset shows the absorber layer is immersed in an alkaline solution (pH \textasciitilde 14) and acidic solution (pH \textasciitilde 1) for 24 hours in the sealed bottles.} 
\end{figure}

The schematic of the synthesis process of the absorber layer is shown in Fig. \ref{fig:absorber}a. The fabrication process is simple and scalable, which is summarized as three steps: hot-pressing, spraying, and drying. In this study, the MF with the dimension of 10 cm $\times$ 6 cm $\times$ 2 cm is selected as an elastic skeleton for immobilization of the Black 3.0 paint spray. MF is a commercially available low-cost polymer material, and its high porosity, low density, excellent hydrophilicity, and elasticity render it highly applicable for the solar steam generation devices \cite{lin2018integrative,li2019scalable}. Hot-pressing treatment of the MF with a compression ratio of 4 has been experimentally confirmed to be advantageous to increase its elasticity and fatigue resistance. More importantly, the compressed three-dimensional network of the foam can dramatically increase its density to effectively increase the ability of water absorption. In laboratory experiments, the pressed MF is cut into a 1 mm-thick circle with a diameter of 47 mm, fitting the size of a 100 ml beaker. 

The inexpensive Black 3.0 paint has gained great attention in creative art and as an superb solar photothermal material. As the blackest and most matte acrylic paint, Black 3.0 paint exhibits nearly perfect solar absorption at up to 99\% of visible light. The paint can be easily applied via brushing or spraying onto most surfaces, such as wood, paper, canvas, and plastic. Last but not the least, the paint is light in density and very suitable to the desalination process.

Figure S2 shows the transmittance spectrum of the Black 3.0 paint hydraulic pressed with KBr powder showing corresponding wavenumber of functional groups. Then the diluted paint is sprayed on the pressed MF sheet using a spray gun. The distance between the spray head and the MF sheet is about 25 cm. Considering the porous surface of the pressed MF sheet surface, it turns out that a couple of thin coating layers are better than one the thick layer. Between each coating process, drying it thoroughly with hot air is necessary at 190$^\circ$C for 5 minutes. After thoroughly drying, the Black 3.0 paint-coated MF sheet is rinsed in deionized (DI) water for several times to remove the residual impurities. Hereafter, keep this sheet in the oven at 60$^\circ$C for drying and then use it as the solar absorber layer in the evaporation device. The absorber layer can be easily prepared requiring no complex process or expensive equipment, and therefore, most suitable for volume production. As shown in Fig. \ref{fig:absorber}b, the absorber sheet exhibits a superb absorption ranging from 95\% to 99\% at the wavelength from 0.2 $\mu$m to 2.5 $\mu$m, indicating that it acts as an efficient broadband solar radiation absorber, whereas the pure pressed MF shows poor absorptivity. Besides, this absorber also exhibits an angular-independent hemispherical absorptivity (Fig. S1). Figure \ref{fig:absorber}c shows the surface temperature of the absorber sheet in the air under 1 sun for an hour. The time-dependent temperature changes and images are captured by an infrared camera at room temperature, displaying the maximum temperature of the sheet over time. In taking the temperature measurement, the tested absorber sample is placed on a PVC foam plate to minimize the heat exchange with the base below. Upon light illumination, the surface temperature of the absorber sheet rises sharply to an equilibrium temperature around 100$^\circ$C within the initial 2.5 minutes and then floats slightly around this temperature afterward, indicating an excellent photothermal performance of the absorber sheet. The inset of Fig. \ref{fig:absorber}c demonstrates the change in surface temperature more clearly in the initial 4 minutes.

Figures \ref{fig:test_absorber}a - \ref{fig:test_absorber}d show the microscopic structures of the absorber. The pristine MF has polygon networks with a size of about 200 $\mu$m and the diameter of the fibers is around 7 $\mu$m (Fig. \ref{fig:test_absorber}a). While the geometric shape of the networks collapses with the reduced network size (Fig. \ref{fig:test_absorber}b), it retains its porous structure, enabling the transportation of water. One-layer Black 3.0 paint coating forms the beads along the fibers (Fig. \ref{fig:test_absorber}c), and the three-layer coating causes an increase in the fiber diameter from 7 $\mu$m to 50 $\mu$m (Fig. \ref{fig:test_absorber}d) and fully-covered Black 3.0 paint coated MF presents nearly unity absorptivity in solar irradiance region. The fully covered MF still keeps a highly-porous structure facilitating the water transport and vapor release, which results in the high water evaporation rate. To evaluate the stability of the absorber sheet, the absorber layers undergo severe tests, kept in the boiling water (around 90$^\circ$C) for 1 hours and immersed in the alkaline solution (pH \textasciitilde 14) and acidic solution (pH \textasciitilde 1) for 24 hours in the closed bottles. After this harsh thermal and chemical stability test, no obvious changes in the appearance are observed and their absorption spectra are basically consistent with the original one (Fig. \ref{fig:test_absorber}e).

\begin{figure}[!t]
\centering
\includegraphics[width=0.8\textwidth]{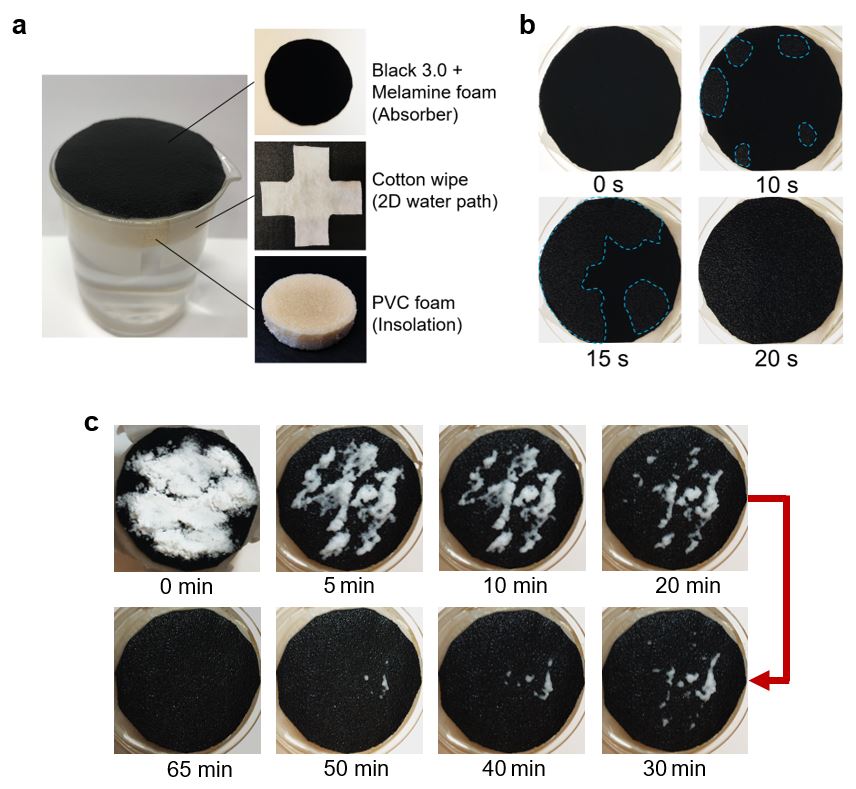}
\caption{ \label{fig:device_wet_salt} The self-floating evaporation device with surface wettability and salt rejection ability tests. (a) Photographs of the three-layer evaporation device placed in a beaker filled with freshwater. (b) Wetting process of the evaporation device placed on the surface of the water. The blue-dash enclosed regions exhibit the wetting areas. (c) The salt rejection progress of the evaporation device. The evaporation device is placed in a beaker filled with 3.5 wt$\%$ NaCl solution, and NaCl is originally stacked on the top surface of the absorber.} 
\end{figure}

Localizing the heat converted from solar energy to the air/water interface is the critical step to achieve high evaporation performance. 
Figure \ref{fig:device_wet_salt}a demonstrates a self-floating evaporation device structure which is composed of three layers. The top layer is a solar absorber layer made of the MF sheet combined with Black 3.0 paint and it plays an essential role in an efficient absorption of the solar radiation. Beneath the absorber layer is a two-dimensional water path enabled by a cotton wipe that is cut into the shape of a circle with 4 extended strips. The round area of the cotton wipe is kept the same as the absorber area, and the bottom edges of 4 strips are in direct contact with the bulk water, which decreased the contact area between the evaporation surface and the bulk water. The cotton wipe is used to transport water due to its strong capillary force and to reduce the direct contact between the bulk water and the absorber layer. Simultaneously ensuring the efficient water supply to the solar heating area and minimizing the heat dissipation to the bulk water is crucial to achieve the high-efficiency interfacial evaporation. The interlocked fibers of cotton wipe do not come apart when soaked in water for a long time, that contributes to the durability and stability of the device. In the working process, the cotton wipe pumps the bulk water towards the heating surface through the 4 strips around the floating foam relying on its strong capillary wicking effect. The bottom layer is the thermal insulation wrapped with the cotton water path. The thermal insulation is made from PVC foam with only closed pores which are impermeable to water, and its low thermal conductivity (\textasciitilde 0.03 W m$^{-1}$ K$^{-1}$) almost reduces the downward thermal dissipation from the evaporation surface. Thanks to the low density of the thermal insulation and the simple structure of the whole device, the evaporation device can be placed on the water and move together with the waving water surface achieving self-floating for continuous operation. 

The surface wettability of the evaporation device also plays an important role in the steam generation process. As shown in  Fig. \ref{fig:device_wet_salt}b, the evaporation device is placed in a beaker filled with DI water at the initial temperature of 21$^\circ$C, and its wetting process is recorded from the top view. It is obvious that the water immediately reaches the surface of the absorber layer from the regional edge (the enclosed areas in the blue dashed), and then the wet area quickly expands to the central area until it covers the entire evaporation surface. The surface wettability owes much to forceful hydrophilicity of the two-dimensional water path of the cotton and the porous structure of the absorber layer, which assures ample water supply to the evaporation surface.

Avoiding salt accumulating in the evaporation device remains a significant character for a self-floating solar evaporation structure utilizing heat localization \cite{ni2018salt}. Figure \ref{fig:device_wet_salt}c shows the progression of salt dissolution which demonstrates the salt rejection ability of the three-layer device. In this experiment, a three-layer structure with a diameter of 47 mm floated in a 100 ml beaker of 132g 3.5 wt\% NaCl solution with 21$^\circ$C initial temperature, and 1.7g of additional solid NaCl is placed directly on the top of the absorber surface. Upon contact with water, the solid NaCl on top starts to dissolve due to the movement and exchange of solution inside the absorber layer and two-dimensional water path between the device surface and the bulk water below the insulation. After approximately 65 minutes, the three-layer device fully rejects the salt, that indicates its good salt rejection ability.

The steam generation ability of the evaporation device at a laboratory scale is tested combining an irradiation system to simulate the solar radiation, a weighing system monitoring the mass change of water in a beaker during a certain period of irradiation, and an infrared camera monitoring the real-time temperature. Both 3.5 wt\% NaCl solution and DI water are prepared to keep the initial temperature at 21$^\circ$C in the 100 ml beakers. The evaporation device floated on water in the beaker filled with 127g of water, beneath the beaker is an electronic balance connected to the computer, and the simulated solar radiation was provided by a solar simulator. When the illumination began, the solar steam generation is determined by the balance, and the real-time temperature of the absorber sheet and water surface are monitored by an infrared camera. The laboratory temperature and humidity are kept at about 22$^\circ$C and 21\%, respectively.

\begin{figure}[!t]
\centering
\includegraphics[width=0.8\textwidth]{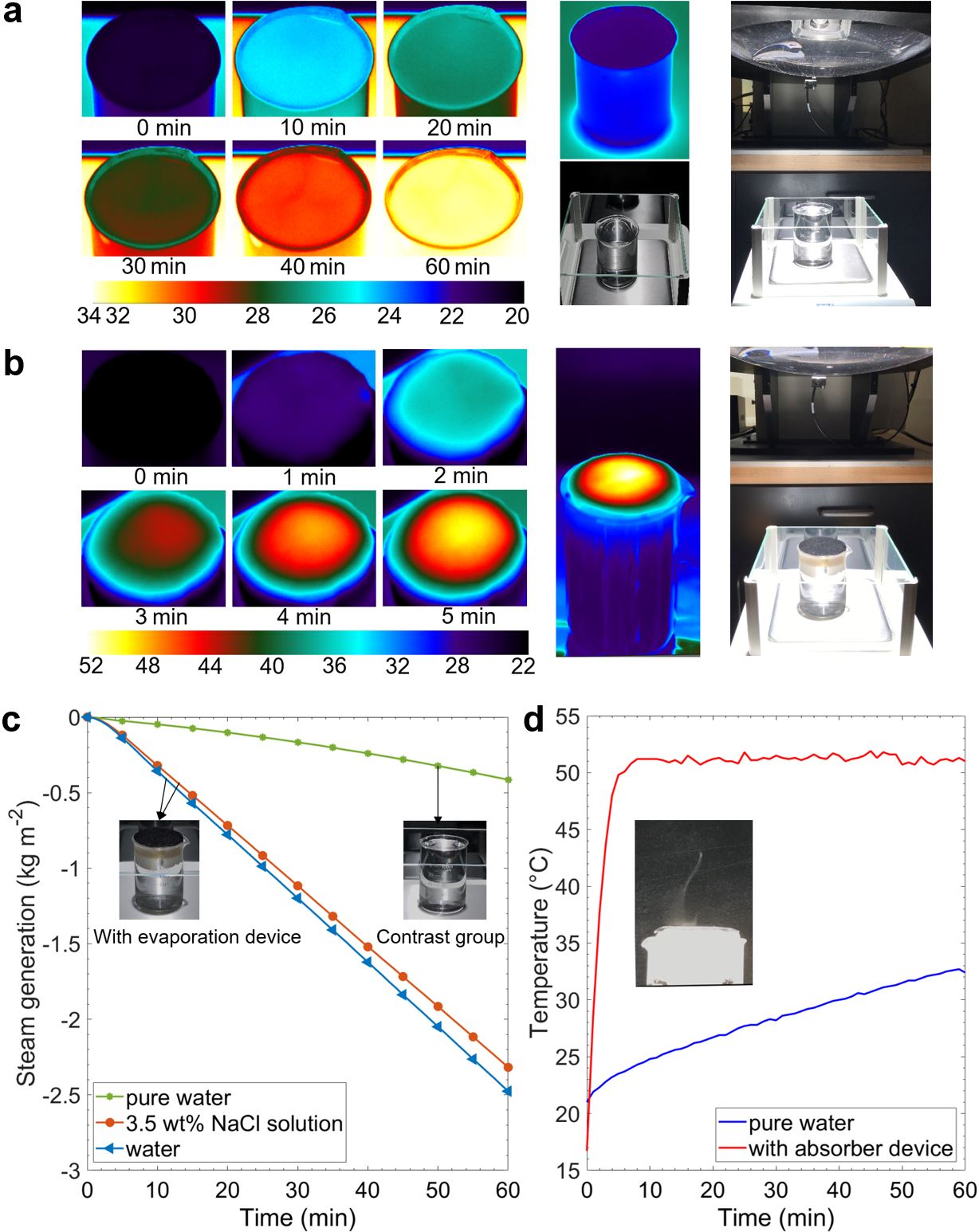}
\caption{ \label{fig:lab_experiment} (a) Left: Infrared (IR) thermal images of the top surface of a water-only beaker under 1 sun illumination over exposure time. Middle: IR thermal image and the photograph of the beaker filled with DI water only under 1 sun after one hour. Right: The photograph of the beaker filled with water only placed on an electrical balance under 1 sun illumination. (b) Left: IR thermal images of the evaporation device under 1 sun illumination in the initial 5 minutes. Middle: IR thermal image of the evaporation device floated on water in the beaker under 1 sun illumination after one hour. Right: The photograph of the beaker with the evaporation device placed on an electrical balance under 1 sun illumination. (c) The change of mass of water over time of DI water only and with the evaporation device placed at the interfaces of the DI water, and 3.5 wt\% NaCl solution. (d) The maximum temperature profiles of the pure water surface and the absorber layer under 1 sun illumination over exposure time. The inset is a visible steam flow generated under 1 sun irradiation.} 
\end{figure}

Figure \ref{fig:lab_experiment} compares the surface temperatures of pure water and 3.5 wt\% NaCl solution with and without the evaporation device under 1 sun illumination. The temperature data and images are captured by an infrared camera. Figure \ref{fig:lab_experiment}a shows the surface temperature distribution of the beaker filled with water without the evaporation device over the irradiation time. The surface temperature rises slowly under irradiation owing to the poor light harvesting efficiency and its maximum temperature profile shown in Fig. \ref{fig:lab_experiment}d slopes gently from the initial temperature of 21$^\circ$C to 32$^\circ$C for an hour. In contrast, upon light illumination, the temperature of the absorber layer rises sharply to an equilibrium temperature around 50$^\circ$C, indicating good photothermal performance. As shown in Fig. \ref{fig:lab_experiment}c, for the water only experiment, the steam generation rate is measured to be 0.41 kg m$^{-2}$ h$^{-1}$ under 1 sun irradiation (1 kW m$^{-2}$). In contrast, the steam generation rate of evaporation device in DI water reaches up to 2.48 kg m$^{-2}$ h$^{-1}$ under the same experimental conditions, 6.04 times higher than that of pure water. Most significantly, when the evaporation device floats on 3.5 wt\% NaCl solution, it sports a comparable steam generation rate of 2.32 kg m$^{-2}$ h$^{-1}$, 5.65 times higher than that of pure water. Evaporation efficiency, $\eta_{evap}$ = $\Dot{m}h_{fg}$/$Q_{s}$, is an important parameter to evaluate the steam generation performance, where $\Dot{m}$ is the water evaporation rate (kg m$^{-2}$ h$^{-1}$), $Q_{s}$ is the power density of irradiation (kW m$^{-2}$), and $h_{fg}$ is the total enthalpy of vaporization of water (kJ kg$^{-1}$). In particular, $h_{fg}$ is regarded as the sum of the sensible heat (121.26 kJ kg$^{-1}$) and the temperature-dependent enthalpy of vaporization (2382.7 kJ kg$^{-1}$) in this text. Thus, we obtain $\eta_{evap}$ = 172.5\% under 1 sun illumination. More detailed calculation of the evaporation efficiency and the photothermal efficiency are provided in the supplementary materials.


\begin{figure}[ht]
\centering
\includegraphics[width=1\textwidth]{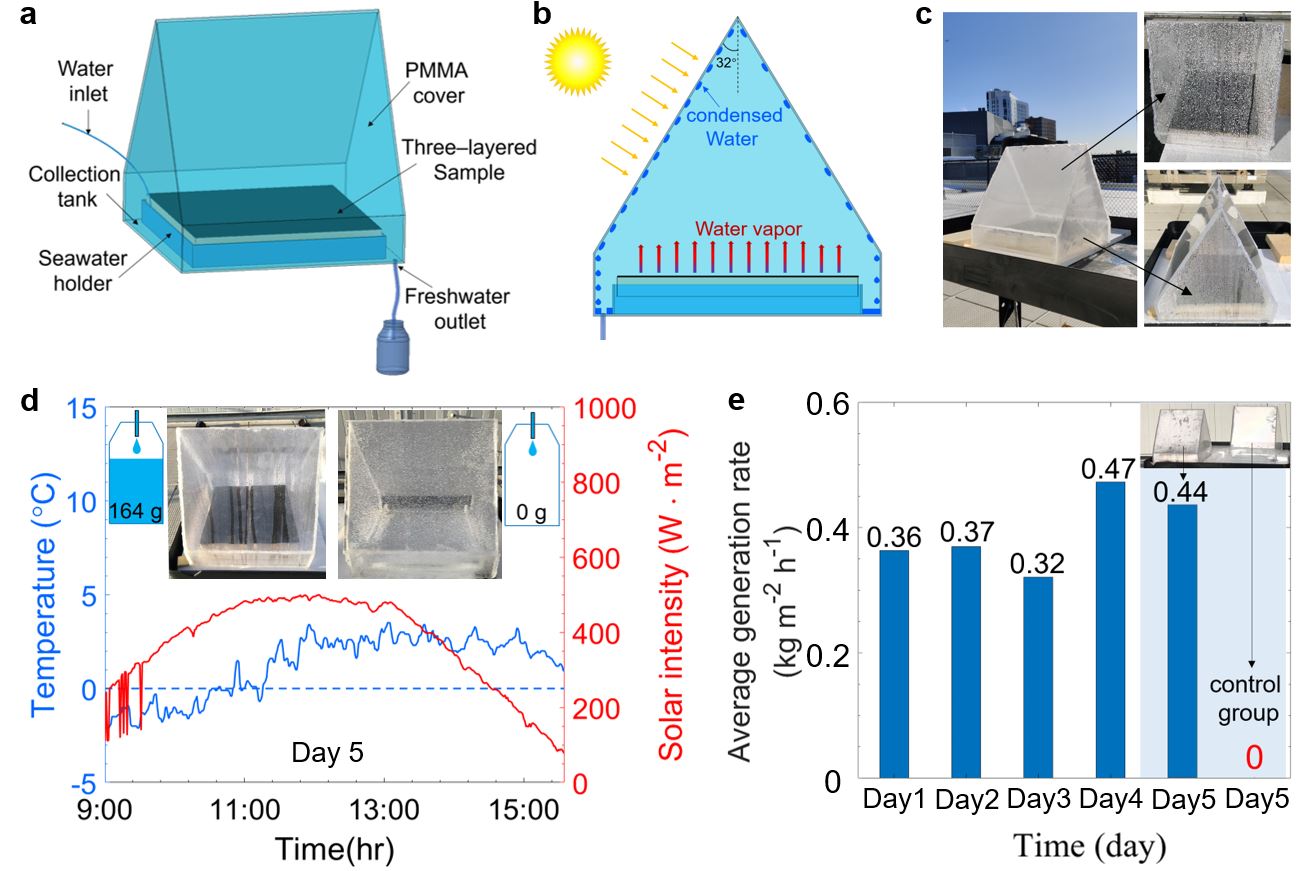}
\caption{ \label{fig:outdoor_1} Rooftop experiments with the evaporation system under natural sunlight in January 2020 in Boston, MA, USA. (a) Three-dimensional model of the evaporation system with a PMMA condensation cover in the dimension of 30 cm $\times$ 30 cm for desalination. (b) Schematic illustration of the solar desalination system. (c) Photograph of the evaporation system placed on a utility cart (left) and the dense condensed water droplets on the interior surface of the condensation cover and sidewall (right) during the experiment. (d) The contrast experiment with an evaporation device (left inset) and without (right inset) on the Day 5. The blue and red curves show the change of ambient temperature (left y-axis) and the solar intensity (right y-axis). Two bottles inset show the weight of the collected water from the contrast experiment. (e) The averaged outdoor steam generation rates of the 3.5 wt\% NaCl solutions for 5 days in January 2020. } 
\end{figure}

\section{Field Test}
We have performed field test of this evaporation device to demonstrate the feasibility of scalable production of freshwater. We constructed a prototype of the evaporation system with a condensation chamber and an evaporation device of a large area of the absorber (24 cm $\times$ 24 cm ) inside, as shown in Figs. \ref{fig:outdoor_1}a and S3. The condensation chamber made of Polymethyl methacrylate (PMMA) board surrounds the entire evaporation device to capture condensate of the evaporated steam. According to the transmission spectrum of the PMMA board, shown in  Fig. S4, the condensation structure is well transparent in the solar spectrum to allow the solar irradiance to reach the evaporation device, while it is opaque in the mid-infrared region to confine the infrared emission of the evaporation device inside the chamber to contribute to maintaining a relatively higher evaporation temperature. As shown in Fig. \ref{fig:outdoor_1}b, the tilted angle of the condensation cover in the dimension of 30 cm $\times$ 30 cm is fixed to be 32$^\circ$ considering the latitude of Boston city, allowing the solar flux to reach the evaporator without refraction. The photograph of Fig. \ref{fig:outdoor_1}c shows the prototype is placed on the polystyrene foam with a low thermal conductivity to reduce the thermal flux between the bottom of the prototype and the utility cart, and the condensed water is observed on the condensation cover (Fig. S5). The condensed water droplets fall to the collection tank and eventually flow to a chemical storage bottle. It has been confirmed that the sodium concentration of the desalinated water is 5 mg l$^{-1}$, that is much lower than the standard of the drinkable water specified by the World Health Organization (WHO) \cite{world2011safe}. The validated prototype is tested on the roof of Snell Engineering Center at Northeastern University, Boston, MA, USA, and water collections are measured over 5 days in January 2020. The specific experiment dates, instantaneous incident sunlight, ambient temperature, wind speed, and humidity are recorded and provided in the supplementary materials. The 3.5 wt\% NaCl solution, which simulates the average salinity of seawater all over the world, is used during the field test and the seawater holder is filled up through the water injection tube after the daily use.

The average evaporation rate of drinkable freshwater, shown in Fig. \ref{fig:outdoor_1}e, varies from 0.32 kg m$^{-2}$ h$^{-1}$ to 0.47 kg m$^{-2}$ h$^{-1}$ under various solar intensity and ambient temperature. The experiment data and detailed weather condition data are provided in Figs. S6 and S7. To demonstrate the promising solar steam generation ability of this evaporation system, the prototype, and the control group, the same condensation chamber without a three-layer evaporation device inside is conducted for a contrast experiment on day 5 (Fig. \ref{fig:outdoor_1}d). At the end of the experiment, 164g of water is collected from the prototype, while the control group collects nearly zero water in the same-day operation. In the rooftop experiments, it is well noted that the weight of collected freshwater refers to the weight of water in the chemical storage bottle, not exactly the weight of condensed water produced. The inset of Fig.  \ref{fig:outdoor_1}d shows the end state of the contrast experiment. Obviously in the control group, dense condensed water droplets can be observed on the cover without forming the water dripping flow (Fig. S8). In day 4, the daily condensate rate (2.83 kg m$^{-2}$ d$^{-1}$) in winter produced from the evaporation device of the area of 0.0576 m$^{-2}$ is comparable to a previous work (2.81 kg m$^{-2}$ d$^{-1}$) in the summer \cite{ni2018salt}, which proves a reliable and scalable production of freshwater using our design in four seasons.

\section{Conclusions}
We have demonstrated a high-performance, low cost, and simple interfacial three-layer steam generation device based on the novel commercial Black 3.0 paint, which is first applied to the solar steam generation as a promising candidate for the photothermal materials. Black 3.0 paint sprayed on a sheet of MF serves as the top solar absorber layer that can widely absorb the solar radiation and efficiently convert it into heat. The absorber layer which is placed on a PVC foam plane under 1 sun irradiation shows a significant temperature gradient and reaches up to 100$^\circ$C of equilibrium temperature. The absorber is shown to sport strong thermal and chemical stabilities under harsh environment, and it exhibits a remarkable salt rejection ability. All these performance contribute to the long-term durable steam generation process. In the laboratory experiments, enabled by the assistance of the low thermal conductivity PVC foam to reduce the heat loss and the cotton wipe to provide sufficient water through two-dimensional water path to the heating region, the evaporation device has reached a striking steam generation rate of 2.48 kg m$^{-2}$ h$^{-1}$, 6 times higher than the natural evaporation, and a highlighted evaporation efficiency of 172.5\%\ under 1 sun irradiation at the room temperature, surpassing most of the reported works. Even in the rooftop experiments during a cloudy winter day in Boston, MA with the maximum environment temperature of 4$^\circ$C, 2.83 kg m$^{-2}$ d$^{-1}$ of water can be collected. This simple, low cost, and easy to manufacture evaporation device is highly beneficial to large scale water desalination and purification applications.

\section{Experimental section}
\subsection{Materials}
The commercial MF is purchased from South Street Designs company (UPC: 089902974060) with the dimension of 10 cm $\times$ 6 cm $\times$ 2 cm (\$0.3/piece). Black 3.0 paint is purchased on the Culture Hustle (\$0.146/ml). Webril pure cotton wipe is in the size of 0.2 m $\times$ 0.2 m (\$2.99/m$^{2}$). The PVC foam insulator sheet is purchased from the McMaster-Carr with the dimension of 813 mm $\times$ 1219 mm $\times$ 13 mm (\$58.7/m$^{2}$). 

\subsection{Sample preparation}
Solar absorber layer is fabricated as follows: pristine MF is thoroughly washed several times with ethanol and deionized (DI) water and then put in an oven kept at 60$^\circ$C in preparation for the hot-pressing treatment. After it is totally dried, the MF is pressed at 200$^\circ$C for 6 minutes with a compression ratio of 4, which is the height ratio of pristine MF to hot-pressed one. Serving as the skeleton of the absorber, hot-pressed MF is cut into the desired shape with a thickness of about 1 mm. Black 3.0 paint is thinned with DI water under vigorous stirring for 5 minutes with a paintbrush, which helps to get a homogeneous mixture. The mass ratio of DI water to Black 3.0 paint is kept in the range of 0.35 $\sim$ 0.4 in the dilution process. Subsequently, the diluted Black 3.0 paint is sprayed onto the MF sheet by a touch-up spray gun (Paasche Airbrush, USA) with a 0.8 mm spray head at the pressure of 70 psi. The distance between the spray head and the MF sheet is about 25 cm. The golden rule is that 3 or 4 thin layers is much better than a single thick layer. And then dry it between each spraying with the hot air blower (Yihua Electronic Equipment Co., Ltd, Guangzhou, China) at a temperature of 190$^\circ$C for 5 minutes. It's worth noting that this drying time is suitable for the specific dimension of the sample in this work. 

A piece of PVC foam (47 mm in diameter and 13 mm in thickness) is utilized as the thermal insulator. Webril pure cotton wipe is cut into a 47 mm circle with four extended strips of 30 mm in width and 20 mm in length. The hydrophilic cotton wipe wrapped around the PVC foam with the four strips tip soaking in the bulk water ensures that the water reaches the upper circular area due to capillary force. Then, the MF sheet is placed over the circular area of the cotton wipe. 

\subsection{Solar steam generation experiments}
The steam generation experiments in the lab are carried out under a solar simulator (Newport, 94081A, class ABB) which supplies solar flux of 1 kW m$^{-2}$ with an optical filter for the standard AM 1.5 G spectrum. 127g of DI water and seawater (3.5 wt\% NaCl) are prepared at the same initial temperature of 21$^\circ$C and placed in the 100 ml beakers with a mouth diameter of 50 mm. The steam generation device is floated on the solution surface and the mass of water is accurately monitored by an electric balance (RADWAG, PS 1000.X2.NTEP) connected to a computer for recording the real-time mass change. The real-time temperature is monitored by an infrared radiation camera (FLIR, A655sc). 

\subsection{Materials characterizations}
The microscope images are obtained by using Metallurgical Microscope (AmScope, ME520TC-18M3) in the darkfield mode. The reflectivity spectra (UV-Visible-Near-infrared range: 200 nm $\sim$ 2500 nm) are measured by the Jasco V770 spectrophotometer at an incident angle of 6$^\circ$ with the ISN-923 60 mm BaSO$_4$ based integrating sphere equipped with PMT and PbS detectors. The reflectivity spectra are normalized by a PTFE based reflectance standard. The transmittance spectra (Mid-infrared region: 2.5 $\mu$m $\sim$ 20 $\mu$m) are measured by the Jasco FTIR 6600 spectrometer at a normal incident angle with reference to the background spectrum of a hydraulic pressed KBr film (20 psi). The Extech EC400 ExStik salinity meter is utilized to characterized the water quality of the collected water samples.


\section*{Acknowledgements}
This project is supported partly by the Soleeva Energy Innovation Award and the National Science Foundation through grant numbers CBET-1941743.







\newpage








\end{document}